\documentclass[letterpaper, traditabstract,letter]{aa}
\usepackage{graphicx}
\usepackage{txfonts}

\def\cc{cm$^{-3}$\space}
\def\s{s$^{-1}$}
\def\kms{km s$^{-1}$\space}
\def\micron{$\mu$m}
\def\microns{$\mu$m\space}

\def\arcsecs{$^{\prime\prime}$\space}

\def\deg{$^{\circ}$\space}

\def\h2{H$_2$}
\def\n2h{N$_2$H$^+$}

\def\13co{$^{13}$CO}
\def\H13CO+{H$^{13}$CO$^+$}
\def\HCO+{HCO$^+$}
\def\c18o{C$^{18}$O}
\def\12co{$^{12}$CO}

\def\cm2{cm$^{-2}$}

\def\c+{C$^+$}

\def\microns{$\mu$m\space}

\begin{document}
\titlerunning{CII sample of  transition clouds }
\authorrunning{ Velusamy, Langer, Pineda, Goldsmith,  Li, Yorke}
 \author{
           T.\,Velusamy \and
      W. D. Langer \and
      J. L.\,Pineda \and
   P. F. Goldsmith \and
            D. Li. \and
        H. W. Yorke }
\offprints{T.\,Velusamy \email{velusamy@jpl.nasa.gov}}
\institute{Jet Propulsion Laboratory, California Institute of Technology, 4800 Oak Grove Drive, Pasadena, CA 91109-8099, USA}
\title{[CII] observations of H$_2$ molecular layers in transition clouds
\thanks{{\it Herschel} is an ESA space observatory with science instruments provided
by European-led Principal Investigator consortia and with important participation from NASA.}}
\date{Received / Accepted }
\abstract {
 We present the first results on the diffuse transition clouds observed in [CII] line emission at 158 \microns (1.9 THz) towards  Galactic longitudes  near 340\deg  (5 LOSs) \& 20\deg (11 LOSs) as part of the HIFI tests and {\bf GOT C+} survey.
 Out of the total 146 [CII] velocity components detected by profile fitting we identify 53 as  diffuse molecular clouds with associated $^{12}$CO  emission but without $^{13}$CO  emission and characterized by A$_{\rm V}$ $<$ 5 mag.
 We estimate the fraction of the [CII] emission in the diffuse HI layer in each cloud and then determine the [CII] emitted from the molecular layers in the cloud.  We show that  the excess [CII] intensities detected in a few clouds is indicative of a thick \h2 layer around the CO core. The wide range of clouds in our sample with thin to thick \h2 layers   suggests that these are  at various evolutionary states characterized by the formation of  \h2  and CO layers from HI and C$^+$, respectively.
  In about 30$\%$  of the clouds the \h2 column densities (``dark gas'') traced by the [CII]  is  50$\%$ or more than  that traced by $^{12}$CO emission. On the average $\sim$ 25$\%$    of the total \h2 in these clouds is in an \h2 layer  which is not traced by CO. We use the HI, [CII], and  $^{12}$CO intensities in each cloud along with simple chemical models  to obtain  constraints on the FUV  fields  and cosmic ray ionization rates.}
\maketitle
\vspace{-0.5cm}
\section{Introduction}
\label{sec:introduction}
The {\it Herschel} key program {\bf GOT C+} ({\bf G}alactic {\bf O}bservations of {\bf T}erahertz C$^+$) is designed to study the  diffuse interstellar medium (ISM) by observing  with the HIFI instrument the [CII] $^{2}P_{3/2}$ $\to$ $^{2}P_{1/2}$ fine structure line emission and absorption at 1.9 THz (158 \micron) over a volume weighted sampling of 500 lines of sight (LOSs) throughout the  Galactic disk.  The {\bf GOT C+} project is described  by Langer et al. (2010a) and the use of [CII] emission to detect diffuse warm ``dark gas'' (\h2 molecular gas not seen by CO observations) by Langer et al. (2010b). C$^+$ is a major ISM coolant, and its 158 \microns  line is an important tracer of the properties of the
diffuse atomic and diffuse molecular gas clouds. The [CII] line thus enables us to trace an important but to date poorly-studied stage in cloud evolution - the transition  clouds going from atomic  to molecular: HI to \h2 and C$^+$ to CI and CO (Snow \& McCall 2006).  These clouds have a large molecular hydrogen fraction in which carbon exists primarily as C$^+$ rather than as CO (Tielens \& Hollenbach 1985; van Dishoeck \& Black, 1988).  Transition clouds
are difficult to study using the standard tracers (HI or CO) but [CII]  can trace this gas.

 There is growing evidence that a substantial amount of interstellar gas exists as molecular \h2, not traced by CO, for example: from Gamma-ray data from EGRET (e.g. Grenier et al. 2005) and Fermi-LAT (e.g. Abdo et al. 2010);  and, the  infrared continuum in diffuse clouds (Reach et al. 1994).  Goldsmith et al.
(2010) detected warm \h2 in emission beyond the CO extent of Taurus.  Wolfire et al. (2010) have modeled the molecular cloud surfaces  to estimate the amount of ``dark gas'' in the form of molecular \h2  in the \h2/C$^+$ layers and find  it contributes about 30$\%$ of the total mass in clouds with total A$_{\rm V}$  $\sim$ 8 mag.  Here, we present direct observational evidence for the \h2/C$^+$  layer in a number of transition clouds through the detection of an excess  [CII] line emission in them.   We use a sample of 53 transition clouds characterized by A$_{\rm V}$ $<$ 5 mag. and the presence of both HI and $^{12}$CO emissions but no $^{13}$CO.  We analyze the observed [CII] intensities combined  with HI and CO data to obtain an   inventory of the total molecular \h2 in different layers in transition clouds and then  constrain the physical conditions by applying simple models for CO formation and photodissociation.

\vspace{-0.5cm}
\section{Observations and data analysis}
The observations reported here were made as part of the HIFI  performance verification  and  priority science phases.    We observed the [CII]  line at
1900.5469\,GHz towards 16 LOSs in the galactic plane with
the HIFI (de Graauw et al. 2010) instrument on  the  {\it Herschel}
Space Observatory (Pilbratt et al. 2010).  The [CII] spectra were obtained using the wide band spectrometer (with 0.22 \kms velocity resolution, over 350 \kms range) at band 7b and using integration of 800s to 1800s (with rms of 0.1K to 0.2K on data smoothed to 1 km s$^{-1}$). For each target  we used the Load chop (HPOINT) with a sky reference offset by 2\deg in latitude.  The data were processed in HIPE version 3.0 using the standard pipeline for HIFI. Using  a fringe fitting tool within HIPE we were able to mitigate the standing waves in band 7b (Higgins \& Kooi, 2009) to sufficiently low levels to provide good baselines in the [CII] spectra (Boogert, private communication).  The data presented here are in the Galactic plane at l = 337.8\degr, 343.04\degr, 343.91\degr, 344.78\degr, 345.65\degr, 18.3\degr, 22.6\degr, 23.5\degr \&  24.3\degr; out of the plane at b=0.5\deg  for l =  24.3\degr and b=1\deg at l=22.6\degr \& 24.3\degr; at b=-0.5\degr \& -1\degr at l=18.3\degr \& 23.5\degr. Table 1 summarizes all the observational data used in our analysis.

\begin{figure}
\centering
\includegraphics[scale=0.36,angle=-90]{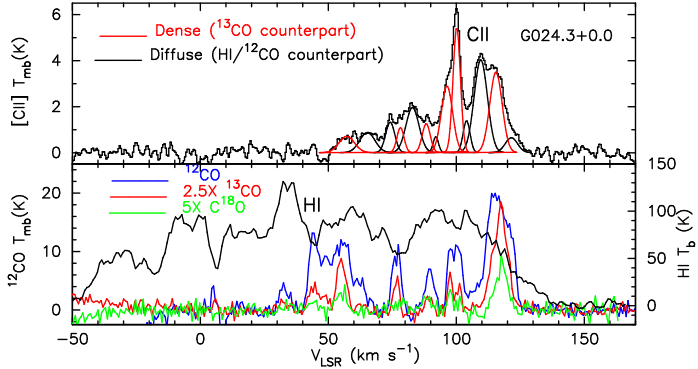}
\caption{   An example of [CII] spectrum for l=24.3\degr, b=0.0\deg and Gaussian fits marked in red and black (top panel) and ancillary data (lower panel).}
\end{figure}

\begin{table}
\caption{Observational Data}
\vspace{-0.25cm}
\begin{tabular}{l c c c c l}
\hline\hline
Line & Survey  & beam & velocity & sensitivity & ref.\\
  & Facility & & chan. & chan.&  \\
  & & & [km s$^{-1}$] & [K km s$^{-1}$] & \\
\hline
[CII] & GOT C+ & 12\arcsecs  & 1.0 & 0.1 - 0.2 & 1,2 \\
 1.9 THz & {\it Herschel} HIFI & & & & \\
\hline
HI & SGPS/ATCA &  132\arcsecs  & 0.84 & 1.6 & 3\\
 &VGPS/VLA& 60\arcsecs  & 0.84 & 2.0 & 4\\
\hline
$^{12}$CO (1-0) & ATNF & 33\arcsecs  & 0.8 & 0.6 & 5\\
$^{13}$CO (1-0) & Mopra &   & 0.8 & 0.1 & \\
C$^{18}$O (1-0) &22-m  &   & 1.6 & 0.1 & \\
\hline
\end{tabular}\\
$^1$This paper; \space\space
$^2$Langer et al. (2010a);  \space\space
$^3$McClure-Griffiths et al. (2005)
$^4$Stil et al. (2006);  \space\space
$^5$Pineda et al. (2010).
\end{table}

\begin{figure}
\centering
\includegraphics[scale=0.3,angle=-90]{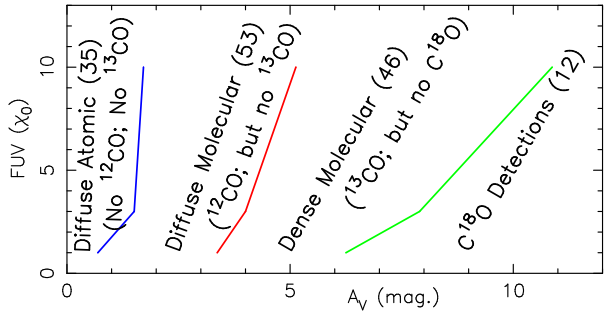}
\caption{ The  [CII] cloud samples identified as a function A$_{\rm V}$ and FUV field.   The lines mark the boundaries  needed to form a detection threshold of CO species. The number of clouds in each category is indicated.  }
\end{figure}

An example of the [CII] spectrum is shown in the top panel in Fig. 1.   The [CII] intensities were corrected for main beam efficiency ($\sim$ 0.63).  For comparison the HI and the CO spectra are shown in the lower panel. The [CII] spectra show many velocity resolved features. All [CII] emission features show an overall  correlation with the HI, though not all HI features show corresponding [CII] emission.  Many [CII] features  are also correlated with CO features.   To separate the individual velocity components  we used multiple Gaussian fitting.  In the case of complex (overlapping) velocity features we used both [CII] and HI profiles together to identify the individual components.  We identified a total of    146  velocity components in all the LOSs.   As seen in Fig. 1 as well as in the examples shown in Langer et al. (2010b) and Pineda et al. (2010) in each spectrum we detect many velocity components. However, their identity as clouds is somewhat uncertain as the  decomposition itself is not very unique and may not be  reliable (e.g. Falgarone et al. 1994).   Though we use the HI profile as an independent check on the features, the  beam sizes (Table 1) are not modeled into the   decomposition.  For simplicity, here we refer to them as clouds, but in reality some of them  may be for example, isolated turbulent clumps, transient fluctuations of larger structures, or superposition of extremely narrow velocity components.  In view of the uncertainties,  for all our quantitative analysis we do not use all of the Gaussian fit parameters. Instead we use the fitted  V$_{LSR}$ to locate a parcel of the gas at a  certain velocity and width.  The I(CII), I(HI), I($^{12}$CO) intensities for each cloud were then obtained, in a consistent manner, by integrating the intensities (T$_{mb}$) over the velocity width ($\Delta$V) centered at the respective V$_{LSR}$ (except in a few cases which are confused by the adjacent component).

 We identified 58 [CII] components as dense molecular clouds traced by their $^{13}$CO emission   (e.g. the red Gaussian fits  in Fig. 1)  and these are discussed in a separate paper by Pineda et al. (2010).   We regard the remaining   88  components without $^{13}$CO counterparts (e.g. the black Gaussian fits in Fig.1) as diffuse clouds, envelopes or transition clouds.
We examined   these  88  diffuse [CII] clouds  by correlating them with the  $^{12}$CO spectra. We found that 53 components have associated $^{12}$CO emission while the remaining 35 have no $^{12}$CO counterparts.  These   35 clouds are labeled diffuse atomic clouds of which 29 are discussed by Langer et al. (2010b).
 To place  our [CII] cloud samples  in the  context of the general interstellar clouds, in Fig. 2 we identify  them   in  an A$_{\rm V}$ - FUV parameter space. We use our 3-$\sigma$   detection limits (Table 1) for [CII], $^{12}$CO, $^{13}$CO, and C$^{18}$O to  estimate the corresponding thresholds of A$_{\rm V}$ and FUV based on the calculations by Visser et al. (2009). The Visser et al. calculations use T$_{gas}$ =100 K, and n$_H$ = 300 \cc similar to what we use below in our analysis.
 Here we present results on 50 transition clouds  excluding 3 for data quality and other issues.

\vspace{-0.5cm}
\section{Results and discussion}

\begin{figure}
\includegraphics[scale=0.35, angle=90]{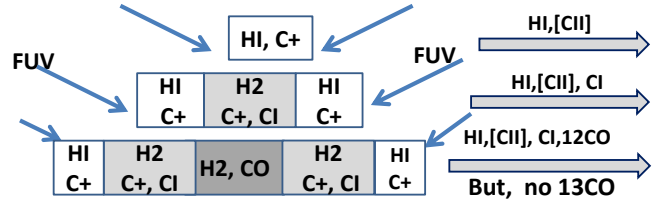}
\caption{  Schematic of the diffuse clouds observed in our sample. }
\end{figure}
\begin{figure}
\centering
\includegraphics[width=0.5\textwidth,angle=0]{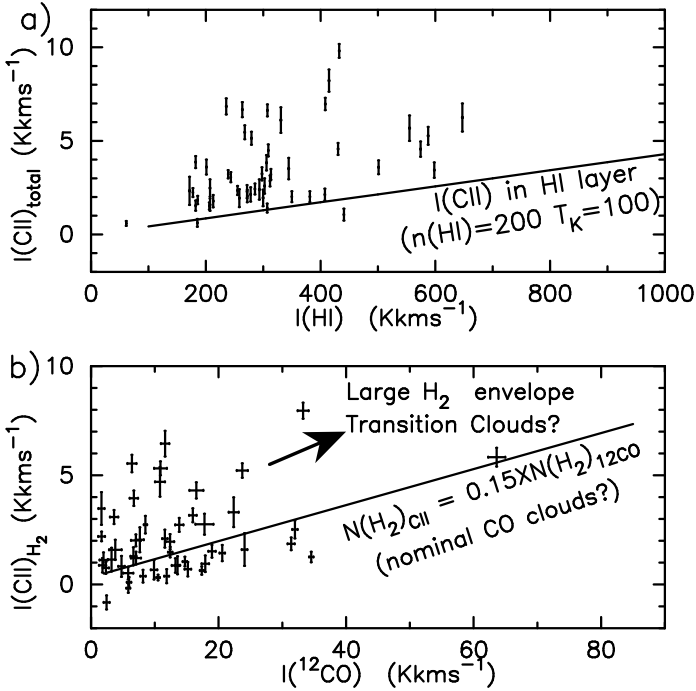}
\caption{  (a)	 The observed [CII] {\it versus} HI intensities: The line is a fit for $I(CII)_{HI}$ in ``nominal''  HI clouds (see text).  The intensities above this line represent that arising from C$^+$ in the \h2 layer surrounding the $^{12}$CO emitting core.
(b)	$I(CII)_{H_2}$, the excess  obtained as I(CII) - $I(CII)_{HI}$ is plotted against I($^{12}$CO).  The line is a fit to [CII]  intensities from ``nominal'' clouds containing about 15$\%$ of the total \h2 in the \h2/C$^+$ layer (see text).     Clouds with  larger \h2 envelopes lie above this line.
  }
\end{figure}

\label{sec:discussion}
\subsection{[CII] sample of transition clouds}
\label{sec:sample}
   In Fig. 2  we find that our [CII] sample of transition clouds are diffuse having A$_{\rm V}$  $\le$ 3 - 4 mag. for reasonable interstellar FUV, in the range of 1 - 10$\chi_D$  (the average FUV intensity  $\chi_D \sim 2.2\times10^{-4}$ erg \cm2 \s sr$^{-1}$ (Draine 1978)).
In Fig. 3 we show a schematic of the diffuse cloud layers.   In the dense cores with $^{13}$CO emission,  the conversion of C$^+$  to CO is more complete while it is partial in these diffuse  transition clouds  due to lack of sufficient self-shielding. All clouds contain some quantity of HI.
 As seen in Fig. 3,   the  observed [CII] emission originates from the purely atomic HI layer along with a contribution from the \h2/C$^+$  layer, while the $^{12}$CO emission originates in the \h2/CO  core. Thus  estimates of \h2 column densities using  $^{12}$CO intensity alone entirely misses the \h2 in the \h2/C$^+$  (``dark gas'') layer.  Therefore,  a complete inventory of molecular \h2 in the cloud requires both the [CII] and $^{12}$CO intensities.
\vspace{-0.5cm}
\subsection{[CII] in the HI/C$^+$ layer}
In Fig. 4a we plot the [CII] intensities against the HI intensities for all 50 transition clouds.    The error bars in Fig. 4a \b  represent the 1-$\sigma$ uncertainties in the respective measured intensities.    In spite of the large scatter we note a lower   bound  to I(CII)   that increases gradually with I(HI) which is consistent with the [CII] emission expected from a HI/C$^+$ layer (Fig. 3).   For quantitative    analysis of [CII] emission from the HI/C$^+$ layer we use the following steps (see also discussion in Langer et al. 2010b):\\
{\bf i)} The observed I(CII) is regarded as the total $I(CII)_{total}$ = $I(CII)_{HI}$ + $I(CII)_{H_2}$,  where $I(CII)_{HI}$ and $I(CII)_{H_2}$ are the emissions originating from the HI/C$^+$  and \h2/C$^+$ layers respectively with no [CII] emission from the  \12co emitting core (Fig. 3).\\
{ \bf ii)} Use the HI intensity, I(HI) to estimate the HI column density,   N(HI) = $1.82\times10^{18}$I(HI)  \cm2.\\
{\bf  iii)} Use N(HI) to estimate the C$^+$ column density in the HI/C$^+$ layer, $N(C^+)_{HI}$   $\sim$ X(C$^+$)N(HI), where  X(C$^+$) = n(C$^+$)/n(HI) is assumed to be 1.5$\times$ 10$^{-4}$.\\
{\bf  iv)} Using the $N(C^+)_{HI}$ above we calculate I(CII)$_{HI}$ $\sim$ f((n(HI),T$_K$) $\times$ $N(C^+)_{HI}$, where the function f accounts for the excitation conditions  for density n(HI) and temperature T$_K$ (see Langer et al. 2010b). Then we can express it in terms I(HI), as    $I(CII)_{HI}$ $\propto$ f(n(HI), T$_K$)I(HI)

We find that on average the form of I(CII) {\it versus} I(HI)
 can be fitted by a straight line obtained for density n(HI) $\sim$ 200 \cc  at temperature T$_K$ $\sim$ 100K as shown in Fig. 4a.   In more massive clouds with HI intensities greater than 1000 K \kms Wolfire et al. (2010) estimate n(HI) $\sim$ 50 -150 \cc and T$_K$ $\sim$ 70 -80 K.  However, all our [CII] clouds are less massive with HI intensities $<$ 600 K km s$^{-1}$.  In the present analysis we assume n(HI) $\sim$ 200 \cc and T$_K$ $\sim$ 100K which seem to describe best the contribution to the [CII] intensity from the HI/C$^+$ layer.
\vspace{-0.5cm}
\subsection{[CII] in the \h2/C$^+$ layer}

Having estimated $I(CII)_{HI}$ arising from the HI/C$^+$ layer we can now calculate the [CII]
excess arising from the  \h2/C$^+$ layer as $I(CII)_{H_2}$ = $I(CII)_{total}$ - $I(CII)_{HI}$.    In  Fig. 4b we show this excess plotted against $I(^{12}CO)$.  Since all the carbon in the $^{12}$CO emitting region is converted to CO we do not expect to see any correlation.  However, in spite of the  large scatter we do note a lower  bound to the excess $I(CII)_{H_2}$   that increases gradually with $I(^{12}CO)$    as seen by the straight line fit in Fig.  4b.   This  suggests the presence of a C$^+$ layer surrounding the $^{12}$CO emitting core as shown in the schematic  in Fig. 3.  The   straight line   in Fig. 4b is   an approximate fit to the lower bound to the   [CII] intensities as a function of $I(^{12}CO)$ and it corresponds roughly to [CII] intensities  for clouds with a \h2/C$^+$ layer containing a \h2 column density $\sim$ 15$\%$ of the   \h2  in the CO core; that is, $N(H_2)_{CII}$  = 0.15$\times$ $N(H_2)_{^{12}CO}$ (see below).   Therefore this line may be regarded as representing the ``nominal'' diffuse CO clouds   which contain a small \h2 envelope around the CO core. However, the clouds with large [CII] excess well above this line  could    represent a sample of clouds in transition with larger \h2 envelopes and relatively smaller CO cores. We can now use this excess
$I(CII)_{H_2}$ and the observed $I(^{12}CO)$  to estimate the \h2 column densities in the \h2/C$^+$ and $^{12}$CO layers respectively.  For $^{12}$CO we use the phenomenological relationship (c.f. Dame et al. 2001):
\begin{equation}
N({\rm H}_2)_{^{12}\rm CO} \sim  1.8 \times 10^{20} I(^{12}\rm CO) \,{\rm cm}^{-2}
\end{equation}
In the \h2/C$^+$ layers the C$^+$ excitation is by \h2 molecules and  we can use the [CII] excess shown in Fig. 4b to derive the $N(H_2)_{CII}$ column density as follows:\\
{\bf  i)} Use the
 $I(CII)_{H_2}$ to calculate the C$^+$ column density, $N(C^+)_{H_2}$  in the \h2/C$^+$ layer,  as a function of density n(\h2) and temperature (T$_K$). Here we assume a higher density of n(\h2) $\sim$ 300 \cc than in the HI layer and a temperature T$_K$ $\sim$ 100K.  \\
  {\bf  ii)} Use this N(C$^+$) column density to  estimate   N(\h2)= N(C$^+$)/2X(C$^+$), where X(C$^+$) = 1.5$\times$ 10$^{-4}$. Thus we get \h2 column density as a function of excess I(CII) for the above assumed n(\h2) and T$_K$ (see Langer at al. 2010b),
\begin{equation}
N({\rm H}_2)_{CII}   \sim  2.8 \times 10^{20}I(CII)_{{\rm H}_2} \,  {\rm
cm}^{-2}
\end{equation}
\begin{table*}[t]
\caption{Selected Transition Cloud Parameters}
\vspace{-0.25cm}
\begin{tabular}{l  c c c c c c c c}
\hline\hline
Cloud &  I(C II) & I(HI)  & I($^{12}$CO) & N(H$_2$) in C$^+$  &N(H$_2$) in $^{12}$CO &  [A$_{\rm V}$ (C$^+$/CO)]$^1$ &\multicolumn{2}{c} {[FUV $\chi_0$]$^2$}  \\
Glong$\pm$latV$\pm$V$_{LSR}$ & Kkms$^{-1}$ &  Kkms$^{-1}$ & Kkms$^{-1}$ & 10$^{20}$\cm2 & 10$^{20}$\cm2 &   mag & $\zeta_{standard}$&40$\zeta_{standard}$ \\
\hline
G343.91+0.00V-14  & 6.2    & 647   & 1.6   & 9.6 & 2.9 &  1.65   &   1.11  &    9.26  \\
G345.65+0.00V-19	&3.2  	& 238    &	1.6 	  &	6.0 &	3.0 &  0.88 &  0.07 &   0.84\\
G337.82+0.00V-127  &  6.1    & 330   &	10.8    &   12.9 & 	19.5 &  1.70 &  1.07&  9.14 \\
G345.65+0.00V-120 &6.7   &	263    &	6.4   &	15.2 &	11.5 &  1.89 &  1.91 &  15.5\\
G345.65+0.00V-114 & 3.9     & 182    & 3.6   & 8.5   & 6.5 &   1.09&   0.15 &  1.61\\
G024.34+0.50V+116 &5.1  &	279 	 &	6.7   &10.1 &	12.1 &  1.43 &  0.46&  4.31\\
G018.26-1.00V+58 & 2.3 &	271    &	1.96  &	  3.1 &	  3.5&  0.59  &   0.02 &  0.33\\
G337.82+0.00V-118  & 8.2   &	414 	 &	11.6  &	17.2 &	20.9 &  1.80&  1.38 &  11.60\\
\hline
\end{tabular}\\
$^1$A$_{\rm V}$ corresponding to the C$^+$/CO layer. \hspace {1cm}
$^2$External FUV radiation field derived for two cosmic ray ionization rates.
\end{table*}

In Fig. 5 we show  the distribution of the ratios of the \h2 column density traced by [CII] to that traced by  $^{12}$CO.
In Table 2 we list a few  diffuse clouds showing a large \h2 layer around the $^{12}$CO emitting core. A majority of the clouds have $N(H_2)_{CII}$  $< $ $N(H_2)_{^{12}CO}$.  In  15 clouds the $N(H_2)_{CII}$  is  50$\%$ or greater than $N(H_2)_{^{12}CO}$.    In this sample of  50 transition  clouds, on   average, $\sim$ 24$\%$ of the total \h2 column density is in the  \h2/C$^+$ layer   which is not traced by $^{12}$CO. Although these estimates are only approximate, they show a likely scenario  in the transition cloud structure.  Lower densities ($\sim$ 100 {cm$^{-3}$) and/or lower temperatures ($\sim$ 50K) will increase the  N(\h2)  in the \h2/C$^+$ layer (required to account for the observed I(CII)) by factors of 2-3, while higher density ($\sim$500 {cm$^{-3}$) will decrease the N(\h2) by a factor of 2.   However, at higher densities  the temperature is likely to be $<$ 100 K and the required N(C$^+$) and N(\h2)  will be larger.

 We can use  N(\h2) in the  \h2/C$^+$ layer and N(HI) in the HI layer derived from  I(CII) and I(HI) to evaluate   A$_{\rm V}$ in the cloud up to the C$^+$/CO transition layer.  We define the C$^+$/CO transition layer as an inner cloud boundary where  X(C$^+$) $\sim$ X(CO) = X(C$_{total}$)/2.  We can now solve for the ratio of external FUV to density  ($\chi$$_0$/n(\h2))  balancing the photodissociation and the CO formation rates at the C$^+$/CO transition layer. We derive analytical photodissociation rates for the attenuation and the self-shielding which are consistent with those given by Lee et al. (1996).  In the warm regions (in all our chemical modeling and analysis we use T$_K$ $\sim$ 100 K) the H$^+$ + OI chemistry dominates CO production over C$^+$ + \h2 (which dominates for T$_K$ $<$ 35 K).  Therefore, for the CO formation rates we use a simple chemical network incorporating the H$^+$ + OI chemistry by extending the approach discussed by Nelson \& Langer (1997) for a CO core surrounded by a warmer tenuous C$^+$ envelope. In our calculation we use the reaction rates given by Glover et al. (2010).  The results for a few clouds are listed in Table 2; the last three columns list   A$_{\rm V}$ up to the C$^+$/CO transition layer and the external FUV, $\chi_0$, in units of Draine radiation field ($\chi_D$).  (Though the solution to the chemical modeling was obtained  as  $\chi_0$/n(\h2) here we give only $\chi_0$ as we have assumed n(\h2)= 300 \cc in our analysis of [CII] intensities). In nearly half of our sample the clouds have very low FUV, $\chi$$_0$ in the range of 0.01 to 0.1$\chi_D$. Such low values seem less likely in the ISM; it has been suggested that the [CII] in the ISM originates from clouds  exposed to FUV,  $\chi_0$ $>$ 10$^{1.2}$$\chi_D$  (Cubick et al. 2008). Using PDR models Pineda et al. (2010)  find that in a [CII] sample of dense molecular clouds the majority have $\chi_0$ = 1 - 10 $\chi_D$. Furthermore it may be noted that   the  H$^+$ + OI chemistry used here is sensitive to the cosmic ray (CR)  ionization rate.  Above we used the standard CR  ionization rate $\zeta_{standard} \sim  2.5\times 10^{-17} s^{-1}$ (c.f. Shaw et al. 2008). However, there is recent evidence of much higher rates in the outer layers (low A$_{\rm V}$) of clouds (Shaw et al. 2008; Indriolo et al. 2007 \& 2009).  We find that using 40$\zeta_{standard}$ increases the derived value of the FUV substantially as shown in the last column in Table 2. At least two of the clouds with high FUV values (G337.82+0.00V-127 and G337.82+0.00V-118) are near the supernova remnant G337.8-0.1, about a shell radius from its boundary at V$_{LSR}$ $\sim$ -122 kms$^{-1}$ (Caswell et al. 1975), and thus may be consistent with our results for higher value for CR ionization.  However, for the cloud G345.65+0.00V-120, though this LOS passes near a HII region (G345.645+0.010), no enhanced radiation feature is observed at this V$_{LSR}$ (Caswell \& Haynes, 1987).

  Our   preliminary  analysis assumes optically thin HI and  C$^+$  emission.  We do not take into account the different beam sizes used in  the  observations.   We do not include the gas traced by CI  in the C$^+$/CO transition zone.     Nevertheless, the results of our simplified approach show a definite statistical trend for the presence of a majority of ``nominal'' diffuse clouds with a thin \h2 layer and a significant fraction of clouds with a thick  \h2 layer   without any accompanying CO.

\begin{figure}
\centering
\includegraphics[scale=0.325,angle=-90]{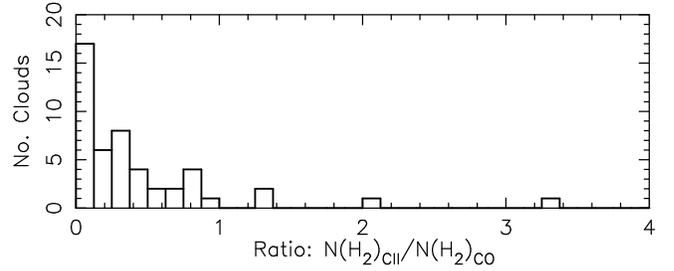}
\caption{ Number  of  transition clouds as a function of the ratio of H$_2$ column densities N(H$_2$ ) traced by  emissions in [CII] to  that in $^{12}$CO.   }
\end{figure}

\vspace{-0.5cm}
\section{Conclusions}
\label{sec:conclusions}
We have observed [CII] line emission in 16 LOSs towards the inner Galaxy and detected   146  velocity resolved [CII] components.  We  identify   53  of these components that are characterized by the presence of both HI and $^{12}$CO but no  $^{13}$CO emission as transition clouds in which the conversion of C$^+$ to CO is partial and a large fraction of carbon exists as C$^+$ mixed with \h2 in a ``dark gas'' layer surrounding the $^{12}$CO  emitting core.    Our results
show that [CII] emission is an excellent  tool to study  transition clouds in the ISM, in particular   as a unique tracer of molecular \h2 which is not easily observed by other means.   In  about 10$\%$ of the clouds the \h2 column density traced by the [CII] emitting layer is  greater than that traced by $^{12}$CO emission. On  average $\sim$ 25$\%$ of the  \h2 in these clouds is in the \h2/C$^+$ layer which is  not traced by CO.  Finally our estimates of the FUV field indicate the   CR ionization is likely much larger than the standard value in the outer layers, consistent with recent determinations from chemical abundances in diffuse regions.

\begin{acknowledgements}
We thank the referee for suggestions.    This work was performed by the Jet Propulsion Laboratory, California
Institute of Technology, under contract with the National Aeronautics
and Space Administration.   The Mopra Telescope is managed by the Australia
Telescope, and funded by the Commonwealth of Australia for
operation as a National Facility by the CSIRO.
\end{acknowledgements}



\end{document}